\documentclass[12pt]{article}
\usepackage[utf8]{inputenc}
\usepackage{amsmath, amssymb, amsthm, bbm, mathrsfs}
\usepackage{geometry}
\usepackage{natbib}
\usepackage{verbatim}
\usepackage{multirow}
\usepackage{booktabs, array, threeparttable}
\usepackage{threeparttable}
\usepackage{rotating}
\usepackage{graphicx}
\usepackage{subfigure}
\usepackage{float}
\usepackage{caption}
\usepackage{amsfonts}
\usepackage{xr}
\usepackage{color}
\usepackage{enumerate}
\usepackage{epstopdf}
\usepackage{threeparttable}
\usepackage{siunitx}
\usepackage{multirow}
\usepackage[colorlinks=true, allcolors=blue]{hyperref}
\usepackage{makecell}
\allowdisplaybreaks
\usepackage[linesnumbered,ruled]{algorithm2e}
\SetKwInput{KwInput}{Input}                
\SetKwInput{KwOutput}{Output}  
\SetKwInput{KwSet}{Set}            
\SetKwProg{Fn}{Function}{:}{}
\usepackage{geometry}
\graphicspath{{./art/}}

\newcommand{\rank}[1]{\text{rank}{\big\left(#1\big\right)}}

\newcommand{\commenting}[1]{}

\newtheorem{theorem}{Theorem}

\newtheorem{remark}{Remark}

\newtheorem{example}{Example}
\newtheorem{condition}{Condition}
\newtheorem{proposition}{Proposition}

\def\sp{\mathcal{C}}
\def\rank{\textrm{rank}}

\def\R{\mathcal{R}}

\title{Power Enhancement of Permutation-Augmented Partial-Correlation Tests via Fixed-Row Permutations}
\date{}
\author{Tianyi Wang, Guanghui Wang, Zhaojun Wang and Changliang Zou\\
{\small\it School of Statistics and Data Science, Nankai University, China}}

\begin{document}

\maketitle

\begin{abstract}
\baselineskip 20pt
Permutation-based partial-correlation tests guarantee finite-sample Type I error control under any fixed design and exchangeable noise, yet their power can collapse when the permutation-augmented design aligns too closely with the covariate of interest. We remedy this by fixing a design-driven subset of rows and permuting only the remainder. The fixed rows are chosen by a greedy algorithm that maximizes a lower bound on power. This strategy reduces covariate-permutation collinearity while preserving worst-case Type I error control. Simulations confirm that this refinement maintains nominal size and delivers substantial power gains over original unrestricted permutations, especially in high-collinearity regimes.

\end{abstract}

\section{Introduction}


We consider a fixed-design linear model of the form $Y=X\beta+Z\theta+\epsilon$, where $(X,Z)\in\mathbb{R}^n\times\mathbb{R}^{n\times p}$ is the design matrix, $\epsilon\in\mathbb{R}^n$ is the noise vector, and $Y\in\mathbb{R}^n$ is the response vector. The objective is to evaluate the partial correlation between $Y$ and $X$ by testing the null hypothesis $\mathcal{H}_0:\beta=0$ against the alternative hypothesis $\mathcal{H}_1:\beta\neq 0$.

Classical F- and t-tests provide valid inference under the assumption of normal noises or in large samples. Permutation and rank-based tests relax the normality assumption, offering validity in finite or large samples under specific noise structures, such as symmetry \citep{hartigan1970exact} and rotational invariance \citep{meinshausen2015group}. A comprehensive review of early developments in this area is presented in Section A of \cite{lei2021assumptionsup}.

Recent advancements have focused on permutation-based tests that maintain finite-sample validity for any fixed design matrix and arbitrary \textit{exchangeable} noise, enhancing robustness across diverse noise distributions. A summary of key examples is provided in Table 1 of \cite{guan2024conformal}. Notably, the Cyclic Permutation Test (CPT) \citep{lei2021assumption} uses a pool of linear statistics of the response vector and ensures that their joint distribution is invariant under cyclic permutations, but it only maintains the nominal significance level $\alpha\in(0,1)$ when $n > (1/\alpha - 1)p$. \cite{wen2024residual} proposed the Residual Permutation Test (RPT) which projects residuals onto subspaces orthogonal to both the original and permuted design matrices, ensuring valid size control when $n > 2p$; however, by replacing unknown noise-projection thresholds with computable upper bounds, its p-value is slightly inflated, rendering RPT mildly conservative in finite samples. In contrast, the Permutation-Augmented Linear Model Regression Test (PALMRT) \citep{guan2024conformal} pairs test statistics from both the original and permuted regression models, guaranteeing worst-case coverage of $2\alpha$ when $n > 2p$, and it empirically controls Type I error.

\subsection{Trivial Power Due to Collinearity}

To illustrate the concept, we first consider a specific instance of PALMRT using a pair of F-statistics. Let $H_M\in\mathbb{R}^{n\times n}$ be the projection matrix onto the column space $\sp(M)$ of $M$. For a set of $n\times n$ permutation matrices $\{P_{\pi_i}\}_{i=1}^B$, each pair of F-statistics is given by:
\begin{align}\label{eq:F}
\begin{aligned}
    T_{i,0}&=\|(I-H_{X,Z,P_{\pi_i}Z})Y\|_2^2=\|(I-H_{X,Z,P_{\pi_i}Z})\epsilon\|_2^2,\\
    T_{0,i}&=\|(I-H_{P_{\pi_i}X,Z,P_{\pi_i}Z})Y\|_2^2=\|(I-H_{P_{\pi_i}X,Z,P_{\pi_i}Z})(X\beta+\epsilon)\|_2^2.
\end{aligned}
\end{align}
A larger difference, with $T_{i,0}<T_{0,i}$, provides stronger evidence against $\mathcal{H}_0$. PALMRT uses the following conformal p-value:
\begin{align}\label{eq:pval}
    p_{\text{val}}=\frac{1}{B+1}\left[1+\sum_{i=1}^B \mathbb{I}(T_{i,0}>T_{0,i})+\frac{1}{2}\mathbb{I}(T_{i,0}=T_{0,i})\right],
\end{align}
which ensures that $pr(p_{\text{val}}\leq\alpha\mid\mathcal{H}_0)<2\alpha$ for any fixed design and arbitrary exchangeable noise, as shown in \cite{guan2024conformal}.

\begin{figure}[!ht]
    \centering
    \includegraphics[width=0.6\textwidth]{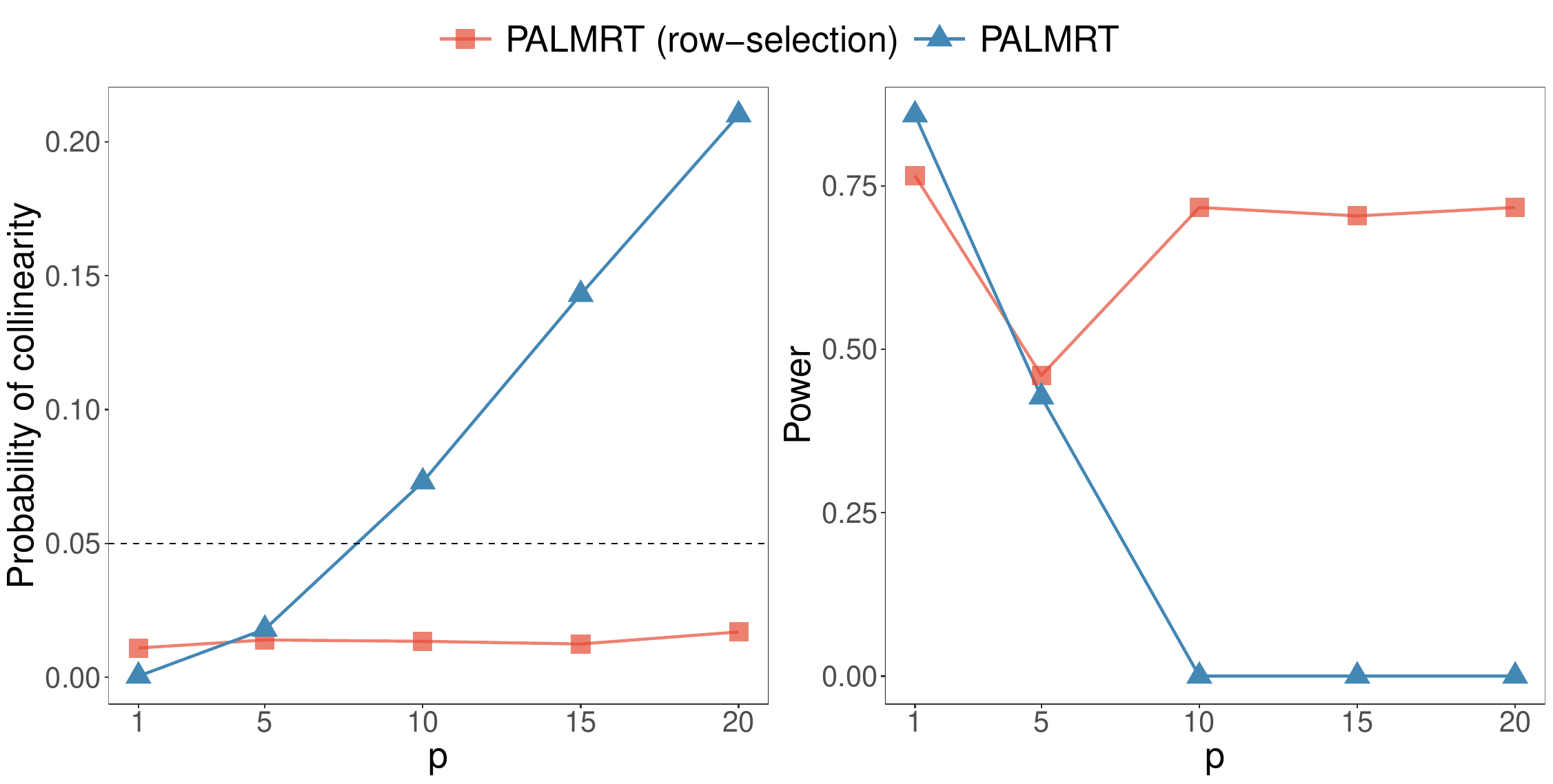}
    \caption{\it \protect \small Comparison of PALMRT's performance with and without row-selection in a paired design with standard Gaussian noise ($Y=4X+\epsilon$) for varying $p$. The left panel illustrates the probability of collinearity, while the right panel shows the test power. The significance level is set to $5\%$.}
    \label{fig:paired}
\end{figure}

However, permutation-augmented spaces for the projections of $Y$ can introduce collinearity. Specifically, when $X\in\sp(P_{\pi_i}X,Z,P_{\pi_i}Z)$, we have $T_{i,0}\geq T_{0,i}$ regardless of the value of $\beta$, failing to reject $\mathcal{H}_0$ when $\beta\neq 0$. This \textit{(near-)collinearity} issue is also noted in \citet{guan2024conformal} (see Section 9, Discussions). To further clarify this issue, consider the following example:

\begin{example}[Paired Designs]\label{exam:paired}
Consider a paired design:
\begin{align*}
(X,Z)=(X,Z_1,\dots,Z_p)=
\begin{pmatrix}
1 & 0_{1\times p} \\  
0_{p\times 1} & I_{p} \\
1 & 1_{1\times p} \\
0_{(n-p-2)\times 1} & 0_{(n-p-2)\times p}
\end{pmatrix},
\end{align*}
where $a_{n\times p}$ denotes an $n\times p$ matrix with all entries equal to $a$, and $I_{p}$ is the identity matrix of size $p\times p$. Let $P_{\pi_i\mid 1\leftrightarrow(k+1)}$ be any permutation matrix that swaps the first and $(k+1)$-th rows. It follows that $X=P_{\pi_i\mid 1\leftrightarrow(k+1)}(Z_k-X)+Z_k$, which implies $X\in\sp(P_{\pi_i\mid 1\leftrightarrow(k+1)}X,Z_k,P_{\pi_i\mid 1\leftrightarrow(k+1)}Z_k)$. This collinearity becomes more pronounced as $p$ increases. As shown in Figure~\ref{fig:paired}, the probability of collinearity, $pr(X\in\sp(P_{\pi_i}X,Z,P_{\pi_i}Z))$, increases with $p$ and leads to trivial power for large $p$.
\end{example}

\subsection{Our Contribution}\label{sec:contribution}

In this paper, we introduce a refinement of PALMRT that addresses the trivial power issues caused by near-collinearity in permutation-augmented regression. The core idea is to perform random permutations by fixing certain rows of the design matrix (i.e., permuting the rest). Specifically, we use random restricted permutation matrices $P_{\pi_i\mid\R}$, where a subset of row indices $\R$ is fixed (see Section~\ref{sec:restrict}). These rows are selected greedily to maximize the angle between $X$ and the space $\sp(P_{\pi_i\mid\R}X,Z,P_{\pi_i\mid\R}Z)$, thereby mitigating collinearity; see Section~\ref{sec:row_selection}. This row-selection strategy guarantees a lower bound on the test power. Although design-dependent, the resulting procedure retains worst-case coverage of $2\alpha$ at any significance level $\alpha$. This follows because all restricted permutation matrices generate a \textit{subgroup}, and uniformly sampling permutations from this subgroup preserves exchangeability---so the conformal-test validity argument still applies.

In Example~\ref{exam:paired}, an intuitive approach is to select a row set $\R=\{2,\ldots,p+2\}$. This ensures $P_{\pi_i\mid\R}Z=Z$ for any $P_{\pi_i\mid\R}$, and thus $X$ only falls into $\sp(P_{\pi_i\mid\R}X,Z,P_{\pi_i\mid\R}Z)$ when $P_{\pi_i\mid\R}X=X$ (which occurs with probability $1/(n-p-1)$). It is implicitly assumed that $X\notin\sp(Z)$, as otherwise $\beta$ would not be identifiable. Figure~\ref{fig:paired} demonstrates that PALMRT with the proposed row-selection strategy effectively reduces the probability of collinearity across various values of $p$, thereby enhancing the test power. Notably, our row-selection strategy successfully identifies the target row set $\{2,\ldots,p+2\}$ in most cases.

\subsection{Related Literature}

Permutation methods have long offered exact, distribution-free inference, yet only recently has their behavior under random resampling been fully characterized. \cite{hemerik2018exact} showed that exactness is retained for any subgroup. \cite{ramdas2023permutation} pushed this further: a randomization device yields valid p-values for any subset and any sampling distribution. A different line of work demonstrates that using a strategically chosen subgroup can substantially improve power and reduce computation: \cite{koning2024moreb} constructed ``representative'' subgroups that dominate random sampling, and \cite{koning2024morea} showed that tiny, well-selected subgroups can even outperform the full group in multiple-testing settings. 

Our work brings these subgroup ideas to permutation-augmented partial-correlation testing. A key difference from earlier work is that even with full-group sampling, the test can still suffer from trivial power whenever permutation-augmented regressors align too closely with the covariate of interest. Our solution is to restrict permutations to a design-driven subgroup with fixed-rows, preserving finite-sample (worst-case) validity of the conformal p-value.

\section{Methodology}

\subsection{Restricted Permutation Matrices and Subgroup}\label{sec:restrict}

Let $[n]=\{1,\ldots,n\}$. A permutation $\pi:[n]\rightarrow[n]$ corresponds to an $n\times n$ permutation matrix $P_{\pi}=(p_{ij})$ where $p_{ij}=1$ if $\pi(i)=j$ and $p_{ij}=0$ otherwise. Applying $P_{\pi}$ to a design matrix $X$ permutes the rows of $X$ according to $\pi$. For the identity permutation $\pi_0$ (where $\pi_0(i)=i$ for $i\in[n]$), $P_{\pi_0}=I_n$. The set of all $n\times n$ permutation matrices forms a symmetric group $\mathcal{S}_n$ of order $n!$ under matrix multiplication.

Given a subset of row indices $\R\subseteq[n]$, a \textit{restricted} permutation matrix $P_{\pi\mid\R}$ is a permutation matrix $P_{\pi}$ with the restriction that $p_{ii}=1$ for all $i\in\R$. The collection of all such restricted permutation matrices forms a \textit{subgroup} $\mathcal{S}_{n\mid\R}$, a subset of $\mathcal{S}_n$ that is closed under matrix multiplication.

\subsection{Our Procedure}

We describe our procedure for a generalized class of paired statistics $(T_{i,0},T_{0,i})$, including paired F-statistics as a special case (see Eq.(\ref{eq:F})). Specifically, for any permutation matrix $P_{\pi_i}\in\mathcal{S}_n$, define
\begin{align*}
   T_{i,0}=T(P_{\pi_i},P_{\pi_0};X,Z,\epsilon),\quad
   T_{0,i}=T(P_{\pi_0},P_{\pi_i};X,Z,\epsilon),
\end{align*}
where $T(\cdot,\cdot;X,Z,\epsilon)$ is a function that satisfies the transferability Condition \ref{condition:transfer} \citep{guan2024conformal}.
\begin{condition}[Transferability]
For any permutation matrices $P_{\pi_i},P_{\pi_j},P_{\sigma}\in\mathcal{S}_n$,
\begin{align*}
    T(P_{\pi_i},P_{\pi_j};X,Z,P_{\sigma}\epsilon)=T(P_{\sigma}^{-1}P_{\pi_i},P_{\sigma}^{-1}P_{\pi_j};X,Z,\epsilon).
\end{align*}
\label{condition:transfer}
\end{condition}

We begin by fixing a subset of row indices $\R=\R(X,Z)\subseteq[n]$ (see Section~\ref{sec:row_selection}), typically depending on the design $(X,Z)$. It then proceeds by randomly sampling $P_{\pi_1\mid\R},\ldots,P_{\pi_B\mid\R}$ from the collection $\mathcal{S}_{n\mid\R}$. A p-value is computed as in Eq.(\ref{eq:pval}), where the random permutation matrices $P_{\pi_i}$ in $(T_{i,0},T_{0,i})$ are replaced by the restricted counterparts $P_{\pi_i\mid\R}$. Importantly, $\mathcal{S}_{n\mid\R}$ forms a subgroup, regardless of the design-dependent nature of $\R$, underpinning the procedure's finite-sample validity.

\begin{theorem}\label{thm:validity}
Consider any fixed design matrix $(X,Z)$ and assume arbitrary exchangeable noise $\epsilon$. Under Condition \ref{condition:transfer}, for any row set $\R=\R(X,Z)$, 
\[
    pr(p_{\text{val}} \leq \alpha\mid \mathcal{H}_0) < 2\alpha
\]
for all $\alpha > 0$.
\end{theorem}

\begin{remark}\label{rmk:tight}
The worst-case bound of $2\alpha$ is inherited from PALMRT and aligns with the bounds for predictive coverage established for multisplit conformal prediction methods \citep{vovk2018cross,barber2021predictive}.
In certain special designs, this bound could be tightened. For example, consider Example \ref{exam:paired} using paired F-statistics with a selected row set $\R$. If additionally $P_{\pi_i\mid\R}Z=Z$ for all $P_{\pi_i\mid\R}\in\mathcal{S}_{n\mid\R}$ (as typically observed in our simulations), we can show that $pr(p_{\text{val}} \leq \alpha\mid\mathcal{H}_0) \leq \alpha$; {see Appendix~\ref{sec:specific_case} for a proof.}
\end{remark}

\subsection{Greedy Row-Selection Strategy}\label{sec:row_selection}

Theorem \ref{thm:validity} permits any design-dependent selection of fixed rows. We propose a strategy motivated by F-statistics (see Eq.(\ref{eq:F})) that enhances power by reducing collinearity.

\begin{proposition}\label{prop:angle}
Consider fixed designs and assume exchangeable noises. For paired F-statistics $(T_{i0},T_{0i})$ and any candidate row set $\R=\R(X,Z)$, 
\[
    pr(p_{\text{val}} \leq \alpha) \geq pr\big(\max_{P_{\pi}\in\mathcal{S}_n}g(P_{\pi},\epsilon;X,Z) \leq \beta^2\cdot F_{D_\R}^{-1}(\alpha+O(B^{-1/2}(\log B)^{1/2});X,Z)\big)+O(B^{-1}),
\]
where $g$ is a design-dependent function of $P_{\pi}$ and $\epsilon$ (see Eq.(\ref{eq:expression_g}) in the Appendix), $D_\R=\|(I-H_{P_{\pi\mid\R}X,Z,P_{\pi\mid\R}Z})X\|_2^2$ is the squared residual norm obtained by projecting $X$ onto $\sp(P_{\pi\mid\R}X,Z,P_{\pi\mid\R}Z)$, and $F_{D_\R}^{-1}(\alpha;X,Z)$ is the $\alpha$-quantile of $D_\R$.
\end{proposition}

Motivated by Proposition \ref{prop:angle}, we seek the row set $\R$ that maximizes the lower bound on power. Up to permutation sampling errors $O(B^{-1/2}(\log B)^{1/2})$, we maximize $F_{D_\R}^{-1}(\alpha;X,Z)$---the $\alpha$-quantile of the squared residual norm $D_\R$. Maximizing this quantile enlarges the angle between $X$ and the projection space $\sp(P_{\pi\mid\R}X,Z,P_{\pi\mid\R}Z)$, thereby minimizing collinearity.

Exploring every subset $\R\subseteq[n]$ is computationally impractical, so we adopt a greedy strategy that adds one row at a time, each time choosing the candidate that gives the largest quantile. A naive implementation would begin with $\R=\emptyset$. This is unsafe: in Example \ref{exam:paired} (see Section~\ref{sec:contribution}), fixing any single index in the target row set $\{2,\ldots,p+2\}$ already forces $pr(X\in\sp(Z,P_{\pi\mid\R}Z))\geq\alpha$ for large $p$ (see Appendix~\ref{sec:initial_shield} for a proof), and Proposition \ref{prop:rank}(i) then implies $F_{D_\R}^{-1}(\alpha;X,Z)=0$. The procedure would therefore produce only the trivial power lower bound. Thus, we first build an initial shield that keeps $X$ safely outside $\sp(Z,P_{\pi\mid\R}Z)$ before the greedy search begins.

\begin{algorithm}[!ht]
\caption{Greedy row-selection}\label{algo:greedy}
\begin{enumerate}
\setcounter{enumi}{-1}
\item \textbf{Input}: design matrix $(X,Z)$, significance level $\alpha$, budget $T$.\\
\textbf{Output}: fixed-row set $\R$.
\item \textbf{Initial shield.}
\begin{enumerate}[a)]
    \item Set $\R^0\leftarrow\emptyset$.
    \item While $pr(X\in\sp(Z,P_{\pi\mid\R}Z))>\alpha$:
    \begin{enumerate}[i)]
        \item For each $k\not\in\R^0$, estimate $\mu_{\R^0\cup\{k\}}(Z)$.
        \item Let $k^*={\arg\min}_k\mu_{\R^0\cup\{k\}}(Z)$ and update $\R^0\leftarrow\R^0\cup\{k^*\}$.
    \end{enumerate}
\end{enumerate}
\item \textbf{Power-oriented augmentation.}
\begin{enumerate}[a)]
    \item For $t\in[T]$:
    \begin{enumerate}[i)]
        \item For each $k\not\in\R^{t-1}$, estimate $Q^t(k)=F_{D_{\R^{t-1}\cup\{k\}}}^{-1}(\alpha)$.
        \item Let $k^*={\arg\max}_k Q^t(k)$ and $Q^t=Q^t(k^*)$. Update $\R^t\leftarrow\R^{t-1}\cup\{k^*\}$.
    \end{enumerate}
    \item $t^* = \arg\max_{0\le t\le T} Q^t$.
\end{enumerate}
\item \textbf{Return:} $\R=\R^{t^*}$.
\end{enumerate}
\end{algorithm}

\textbf{Stage I (initial shield)}: The idea is to make $Z$ and $P_{\pi\mid\R}Z$ similar, thereby pushing $X$ away from $\sp(Z,P_{\pi\mid\R}Z)$ because of $X\notin\sp(Z)$. Define $\mu_\R(Z)=E\{\rank(Z-P_{\pi\mid\R}Z)\}$. Proposition \ref{prop:rank}(ii) shows that $\mu_\R(Z)$ decreases as $\R$ grows, and therefore we suggest adding rows one-by-one until the safety condition $pr(X\in\sp(Z,P_{\pi\mid\R}Z))<\alpha$ is met. The resulting set is denoted as $\R^0$.

\textbf{Stage II (power-oriented augmentation)}: We perform $T$ greedy iterations to enlarge the separation between $X$ and the space $\sp(P_{\pi\mid\R}X,Z,P_{\pi\mid\R}Z)$, thereby improving power. At iteration $t\in[T]$, for every candidate row $k\not\in\R^{t-1}$, compute the quantile $F_{D_{\R^{t-1}\cup\{k\}}}^{-1}(\alpha;X,Z)$. We then select $k^*$ that maximizes the quantile, updating $\R^{t}=\R^{t-1}\cup\{k^*\}$ and storing this best quantile as $Q^{t}$. After $T$ steps we select the iterate $\R^t$ whose $Q^t$ is the largest among $\{Q^0,Q^1,\ldots,Q^T\}$, ensuring a power lower bound no smaller than that of PALMRT.

\begin{proposition}\label{prop:rank}
For any fixed design $(X,Z)$ and row set $\R$, (i) $pr(X\in\sp(Z,P_{\pi\mid\R}Z))\geq\alpha$ implies $F_{D_\R}^{-1}(\alpha;X,Z)=0$; (ii) $\mu_\R(Z)\geq\mu_{\R\cup\{k\}}(Z)$ for any $k\in[n]\backslash\R$.
\end{proposition}

This monotonicity justifies the greedy construction in Stage I and ensures that each added row strictly improves the collinearity shield.

\begin{remark}\label{rmk:MC}
The rank criterion $\mu_{\R}(Z)$, the safety probability $pr(X\in\sp(Z,P_{\pi\mid\R}Z))$, and the squared residual-norm quantile $F^{-1}_{D_{\R}}(\alpha;X,Z)$ can all be approximated via Monte Carlo simulation of restricted permutation matrices $P_{\pi\mid\R}\in\mathcal{S}_{n\mid\R}$. In particular, verifying $X\in\sp(Z,P_{\pi\mid\R}Z)$ reduces to checking whether $\rank(\sp(Z,P_{\pi\mid\R}Z))=\rank(\sp(X,Z,P_{\pi\mid\R}Z))$.
\end{remark}

The complete procedure is depicted in Algorithm \ref{algo:greedy}.

\section{Simulation Studies}


We compare our fixed-row PALMRT to two baselines---the original PALMRT without row-selection and CPT---evaluating finite-sample size and power. Throughout, all experiments use $n = 100$ observations, a nominal level $\alpha=5\%$, and $B = 2,000$ permutations for PALMRT both with and without row-selection. CPT follows the implementation details recommended by \cite{lei2021assumption}, i.e., a genetic pre-ordering of $10,000$ optimization steps followed by $19$ cyclic permutations. For our greedy search we set a budget of $T = 20$ when $p = 15$ and $T = 30$ when $p = 40$. Quantities such as $F^{-1}_{D_{\R}}(\alpha;X,Z)$, $pr(X\in\sp(Z,P_{\pi\mid\R}Z))$, and $\mu_{\R}(Z)$ are estimated according to Remark \ref{rmk:MC} with $2,000$ restricted permutations.

We consider three representative design matrices that span the spectrum from negligible to severe collinearity:
\begin{itemize}
    \item Gaussian: entries of $(X, Z)$ are i.i.d. $N(0, 1)$ with $p = 15$ covariates, where collinearity is negligible.
    \item Paired: the high-collinearity construction of Example \ref{exam:paired} with $p = 15$.
    \item Mixture: entries are drawn i.i.d. from $0.95 \delta_0 + 0.05 N(0, 1)$ with $p = 40$, creating moderate collinearity, where $\delta_0$ represents the Dirac function at $0$.
\end{itemize}
Noise vectors $\epsilon$ are generated from four distributions---standard Gaussian, $t_3$, standard Cauchy, and a sparse multinomial perturbation ($N(0,I_n)+10^4\times \text{Multinomial}(1;(n^{-1},\ldots,n^{-1}))\times (-1)^{\text{Bernoulli}(0.5)}$)---to cover light- and heavy-tailed as well as discrete noise regimes.

For every design-noise combination, we carry out $50$ Monte Carlo replicates; within each replicate, we generate $2,000$ independent noise vectors to estimate the empirical rejection probability---whether for size or power.

\subsection{Size Control}

\begin{table}[htbp]
\centering
\caption{\it \protect \small Median of empirical sizes across designs and noise distributions, at the nominal level $\alpha=5\%$.}
\label{tab:size}
\begin{tabular}{ccccc}
\toprule
\multirow{2}{*}{Design} & \multirow{2}{*}{Noise} & PALMRT\ & \multirow{2}{*}{PALMRT} & \multirow{2}{*}{CPT} \\
&&(row-selection)&&\\
\midrule
Gaussian & Gaussian    & 2.0  & 1.9  & 5.1  \\
         & t$_3$       & 1.9  & 1.9  & 5.1  \\
         & Cauchy      & 1.8  & 1.9  & 5.0  \\
         & Multinomial & 1.6  & 1.7  & 5.1  \\
\\
Paired   & Gaussian    & 4.8  & 0    & 4.9  \\
         & t$_3$       & 5.0  & 0    & 5.1  \\
         & Cauchy      & 4.9  & 0    & 5.0  \\
         & Multinomial & 4.8  & 0    & 5.0  \\
\\
Mixture  & Gaussian    & 0    & 0    & 4.9  \\
         & t$_3$       & 0.1  & 0    & 5.0  \\
         & Cauchy      & 0.5  & 0.1  & 5.0  \\
         & Multinomial & 1.0  & 0    & 5.0  \\
\bottomrule
\end{tabular}
\end{table}

Under $\mathcal{H}_0$ we set $Y=\epsilon$. Table \ref{tab:size} reports the median of empirical sizes across replicates. All three methods control the nominal level in the Gaussian design, where collinearity is negligible. In the paired design, the original PALMRT collapses to $0$ Type I error due to collinearity; our refined PALMRT with row-selection eliminates this pathology and attains $\approx 5\%$. In the mixture design, both permutation-augmented tests become slightly conservative, yet our procedure still rejects more often than PALMRT, which remains degenerate at $0$. Overall, our procedure inherits PALMRT's finite-sample validity but avoids its ``zero-size'' failure modes.

\subsection{Power Analysis}

\begin{figure}[!ht]
    \centering
    \includegraphics[width=0.9\textwidth]{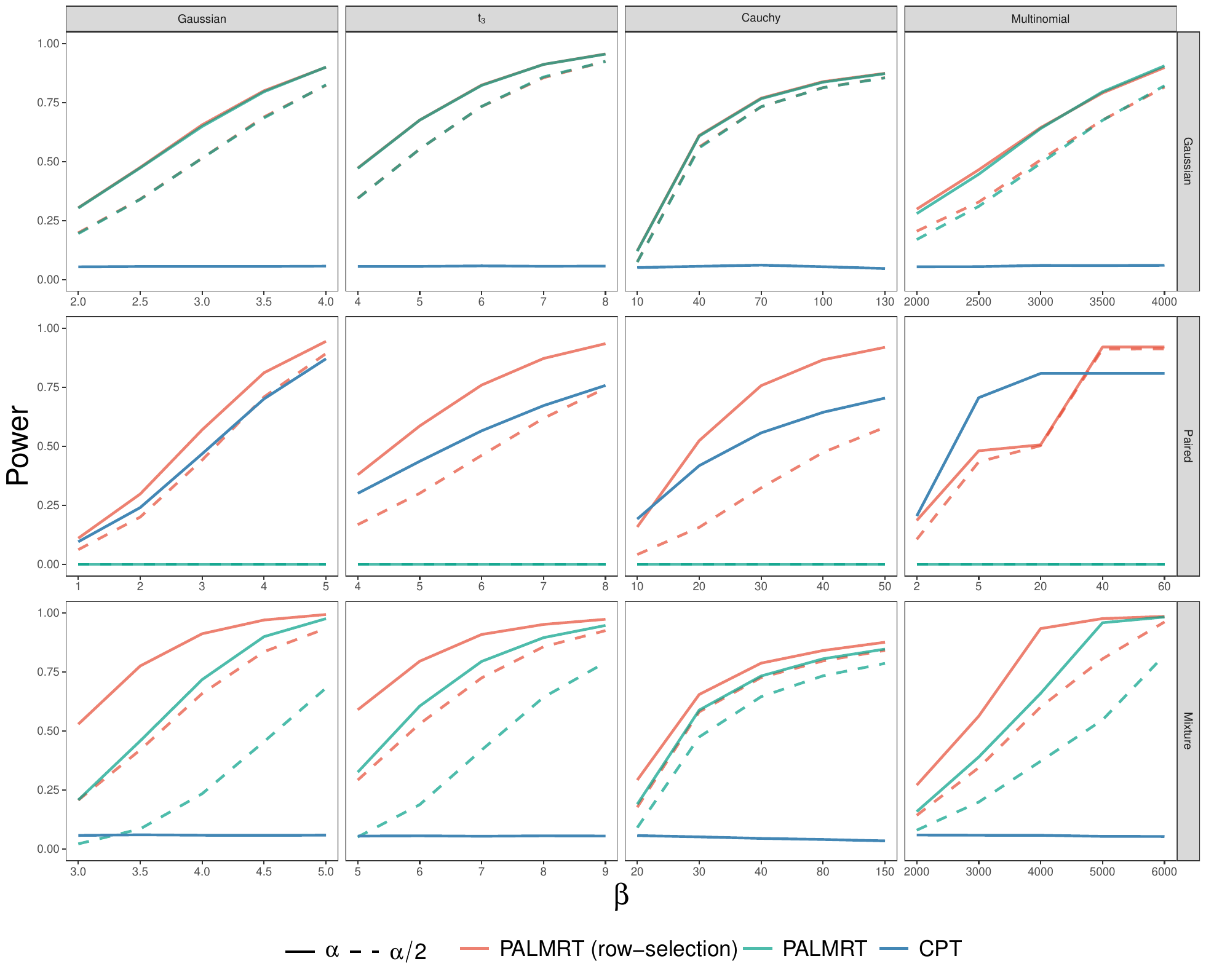}
    \caption{\it \protect \small Empirical power curves versus signal strength at the nominal level $\alpha = 5\%$ across designs and noise distributions. PALMRTs with (red) and without (green) row-selection are each calibrated at $\alpha = 5\%$ (solid lines) and $\alpha/2 = 2.5\%$ (dashed lines), while CPT (blue, solid lines) is shown at $\alpha = 5\%$.}
    \label{fig:power}
\end{figure}

To assess power we generate responses from $Y = X\beta + \epsilon$, varying the signal strength $\beta$. Figure \ref{fig:power} summarizes the median empirical power across replicates. Solid curves correspond to tests run at the nominal level $\alpha=5\%$; dashed lines depicts the row-selected and original PALMRTs calibrated at $\alpha/2 = 2.5\%$, which guarantee worst-case coverage of $2\times(\alpha/2)=5\%$.
\begin{itemize}
    \item Gaussian design: with negligible collinearity, PALMRTs both with and without row-selection perform indistinguishably, confirming that fixing rows does not harm power in benign designs. CPT, by contrast, is effectively powerless because its dimensionality condition $n>(1/\alpha-1)p$ fails when $(n,p,\alpha)=(100,15,5\%)$.
    \item Paired design: the original PALMRT shows trivial power across all noise types. CPT achieves moderate power here because the sparse structure of the design admits a permutation that restores invariance (despite $p=15$). PALMRT with row-selection (at $\alpha=5\%$) dominates CPT in most cases, whereas CPT excels for weak signals under multinomial noises.
    \item Mixture design: the original PALMRT regains power, but our row-selected PALMRT dominates uniformly. CPT again exhibits zero power---now due to both the higher dimensionality ($p=40$) and the denser design.
\end{itemize}

Calibrating both PALMRT variants at $\alpha/2$ predictably lowers power by a small margin, yet our row-selected version still matches or exceeds the original one.

\vspace{0.5cm}
{\small \baselineskip 10pt
\bibliographystyle{asa}
\bibliography{paper-ref}
}
\newpage
\setcounter{equation}{0}
\def\thelemma{S.\arabic{lemma}}
\def\thepro{S.\arabic{pro}}
\def\theequation{S.\arabic{equation}}
\def\thetable{S.\arabic{table}}
\def\thefigure{S.\arabic{figure}}
\renewcommand{\thealgocf}{S.\arabic{algocf}}
\setcounter{lemma}{0}
\setcounter{figure}{0}
\setcounter{table}{0}
\setcounter{algocf}{0}

\def\thesection{A.\arabic{section}}
\setcounter{section}{0}

\noindent{\bf\Large Appendix}

\section{Proof of Theorem~\ref{thm:validity}}

\begin{proof} 
Without loss of generality, label the fixed rows by $\R=\{n-s+1,\dots,n\}$ for some design-dependent integer $s=s(X,Z)$. Let $\R_i=\{n-i+1,\dots,n\}$. Then
\begin{align*}
\mathcal{S}_{n\mid\R_i}=\left\{
\begin{pmatrix}
    P & 0_{(n-i)\times i} \\  
    0_{i\times (n-i)} & I_{i} 
\end{pmatrix}: P\in\mathcal{S}_{n-i} \right\},
\end{align*}
which forms a subgroup of $\mathcal{S}_{n}$. Note that $\mathcal{S}_{n\mid\R_0}= \mathcal{S}_{n}$. By Condition~\ref{condition:transfer}, for each $i\in\{0,\dots,n\}$, $\{T(P_{\pi_l\mid\R_i},P_{\pi_k\mid\R_i},X,Z,\epsilon)\}$ is identically distributed as $\{\tilde{T}_{lk}\}=\{T(\tilde{P}_{\pi_l\mid\R_i},\tilde{P}_{\pi_k\mid\R_i},X,Z,\epsilon)\}$, where $P_{\pi_0\mid\R_i}=I_n$, $\{P_{\pi_l\mid\R_i}\}_{l=1}^B$ and $\{\tilde{P}_{\pi_l\mid\R_i}\}_{l=0}^B$ are randomly sampled from $\mathcal{S}_{n\mid\R_i}$. Thus, this distributional identity property is preserved for $\mathcal{S}_{n\mid\R}$ with design-dependent $s$.
Following similar arguments used in \cite{guan2024conformal}, the collection $\{f(l,\tilde{T})=1+\sum_{k\neq l}(1\{\tilde{T}_{kl}>\tilde{T}_{lk}\})\}_{l=0}^B$ is exchangeable for different $l$, and thus the probability of $p_{\text{val}}\leq \alpha$ can be bounded by the size of $\{l:f(l,\tilde{T}\leq\alpha(B+1))\}$.
\end{proof}

\section{Tight Bound in Paired Designs}\label{sec:specific_case}
\textit{Statement.} If additionally $P_{\pi_i\mid\R}Z=Z$ for all $P_{\pi_i\mid\R}\in\mathcal{S}_{n\mid\R}$, then $pr(p_{\text{val}} \leq \alpha\mid\mathcal{H}_0) \leq \alpha$.
\begin{proof}
    When $P_{\pi_i\mid\R}Z=Z$ for all $P_{\pi\mid\R}\in\mathcal{S}_{n\mid\R}$, we can rewrite $T_{i,0}$ and $T_{0,i}$ as
    \begin{align*}
    T_{i,0}&=\|(I-H_{X,Z,P_{\pi_i\mid\R}Z})Y\|_2^2=\|(I-H_{X,Z})Y\|_2^2,\\
    T_{0,i}&=\|(I-H_{P_{\pi_i\mid\R}X,Z,P_{\pi_i\mid\R}Z})Y\|_2^2=\|(I-H_{P_{\pi_i\mid\R}(X,Z)})Y\|_2^2.
    \end{align*}
    Now we denote $T_{i,0}$,and $T_{0,i}$ as $T_0$ and $T_i$, respectively:
    \begin{align*}
    p_{\text{val}}=\frac{1}{B+1}\left[1+\sum_{i=1}^B \mathbb{I}(T_{0}>T_{i})+\frac{1}{2}\mathbb{I}(T_{0}=T_{i})\right].
    \end{align*}
    According to the transferability condition, we have
    \begin{align*}
        T_{i}=\|(I-H_{X,Z})P_{\pi_i\mid\R}^{-1}Y\|_2^2.
    \end{align*}
    Under $\mathcal{H}_0$, the exchangeability of $\epsilon$ and the same technique used in Theorem 2 of \cite{hemerik2018exact} allow us to conclude that $T_0,\dots,T_B$ are exchangeable. Consequently, by the standard analysis of permutation tests, we have
    \begin{align*}
        pr(p_{\text{val}} \leq \alpha\mid\mathcal{H}_0) \leq \alpha.
    \end{align*}
\end{proof}

\section{Necessity of the Initial Shield}\label{sec:initial_shield}
\textit{Statement.} In paired designs, fixing any single index in the target row set $\{2,\ldots,p+2\}$ already forces $pr(X\in\sp(Z,P_{\pi\mid\R}Z))\geq\alpha$ for $p\geq \alpha(n-1)+1$.
\begin{proof}
    For indices $j,k\in[n]\backslash\R$, define $\mathcal{S}_{n\mid\R,k\rightarrow j}\subseteq \mathcal{S}_{n\mid\R}$ as the subset of permutations that move the $k$th row to the $j$th row while keeping the rows in $\R$ fixed. The cardinality of this subset is given by $\vert \mathcal{S}_{n\mid\R,k\rightarrow j} \vert=(n-\vert \R\vert-1)!$. Furthermore, for any distinct indices $k_1,k_2\in [n]\backslash\R$, $\mathcal{S}_{n\mid\R,k_1\rightarrow j}\cap \mathcal{S}_{n\mid\R,k_2\rightarrow j}=\emptyset$. 
    
    When $\R=\{p+2\}$, note that for $P_{\pi\mid\R,k\rightarrow 1}\in\mathcal{S}_{n\mid\R,k\rightarrow 1}$, we have
    \begin{align*}
        X=P_{\pi\mid\R,k\rightarrow 1}Z_k,\quad \forall k\in\{2,\dots,p+1\},
    \end{align*}
    which implies $X\in\sp(Z,P_{\pi\mid\R,k\rightarrow 1}Z)$. Consequently,
    \begin{align*}
        pr(X\in \sp(Z,P_{\pi\mid\R}Z))\geq p\frac{(n-2)!}{(n-1)!}=\frac{p}{n-1}.
    \end{align*}

    When $\R=\{j\},j=2,\dots,p+1$, we obtain
    \begin{align*}
        X=P_{\pi\mid\R,k\rightarrow 1}(Z_k-Z_j)+Z_j,\quad \forall k\in\{2,\dots,p+1\}\backslash\{j\}.
    \end{align*}
    Similarly, we conclude that $pr(X\in \sp(Z,P_{\pi\mid\R}Z))\geq (p-1)/(n-1).$

    In summary, attempting to fix any single index in the target row set $\{2,\ldots,p+2\}$ yields $pr(X\in\sp(Z,P_{\pi\mid\R}Z))\geq\alpha$, for $p\geq \alpha(n-1)+1$.
\end{proof}

\section{Proof of Proposition~\ref{prop:angle}}\label{sec:power}
\begin{proof}
Denote 
\begin{align}\label{eq:expression_g}
    g(P_{\pi },\epsilon;X,Z)=\Vert(I-H_{X,Z,P_{\pi }Z})\epsilon\Vert_2^2-\Vert(I-H_{P_{\pi }X,Z,P_{\pi }Z})\epsilon\Vert_2^2-2X^T\beta(I-H_{P_{\pi }X,Z,P_{\pi }Z})\epsilon.\tag{A1}
\end{align}
Then
\begin{align*}
    \text{Power}&\geq pr\left\{ \frac{1}{B+1}\left[1+ \sum_{i=1}^B\mathbb{I}(T_{i,0}- T_{0,i}\geq0)\right]<\alpha\right\}\\
    &\geq pr_{\epsilon}\left[ pr_{P_{\pi_i }\in\mathcal{S}_{n\mid\R}} \left( T_{i,0}-T_{0,i}\geq 0\right)<\alpha-O(B^{-1/2}(\log B)^{1/2})\right]-O(B^{-1})\\
    &=pr_{\epsilon}\left\{ pr_{P_{\pi_i }\in\mathcal{S}_{n\mid\R}} \left[g(P_{\pi_i },\epsilon;X,Z)\geq \beta^2D_\R\right]<\alpha-O(B^{-1/2}(\log B)^{1/2})\right\}-O(B^{-1})\\
    &\geq pr_{\epsilon}\left\{ \max_{P_{\pi_i }\in \mathcal{S}_n} g(P_{\pi_i },\epsilon;X,Z)\leq \beta^2F_{D_\R}^{-1}(\alpha-O(B^{-1/2}(\log B)^{1/2});X,Z)\right\}-O(B^{-1}),
\end{align*}
where the second inequality is from the Dvoretzky-Kiefer-Wolfowitz inequality:
    \begin{align*}
        pr\left(\sup_{x\in\mathbb{R}}\left(F_n(x)-F(x)\right)>\varepsilon\right)\leq e^{-2n\varepsilon^2}.
    \end{align*} 
    In the final inequality, we take the maximum on the left-hand side and the corresponding quantile on the right-hand side. The relation $\max_{P_{\pi_i }\in\mathcal{S}_{n\mid\R}}(\cdot)\leq \max_{P_{\pi_i }\in\mathcal{S}_n}(\cdot)$ holds because $\mathcal{S}_{n\mid\R}\subseteq\mathcal{S}_n$.
\end{proof}

\section{Proof of Proposition~\ref{prop:rank}}
\begin{proof}
    (i) Notice that $\sp(Z,P_{\pi\mid\R}Z)\subseteq \sp(P_{\pi\mid\R}X,Z,P_{\pi\mid\R}Z)$ for any fixed design $(X,Z)$ and row set $\R$. Consequently, if $pr(X\in\sp(Z,P_{\pi\mid\R}Z))\geq \alpha$, then
    \begin{align*}
        pr(X\in\sp(P_{\pi\mid\R}X,Z,P_{\pi\mid\R}Z))\geq pr(X\in\sp(Z,P_{\pi\mid\R}Z))\geq \alpha.
    \end{align*}
    Furthermore, when $X\in\sp(P_{\pi\mid\R}X,Z,P_{\pi\mid\R}Z)$, we have $D_{\R}=0$. By the definition of quantile, $pr(X\in\sp(P_{\pi\mid\R}X,Z,P_{\pi\mid\R}Z))\geq \alpha$ implies that $F_{D_\R}^{-1}(\alpha;X,Z)=0$.

    (ii) Without loss of generality, suppose that $\R=\{n-s+1,\dots,n\}$. For $k\in [n]\backslash\R$, note that $\mathcal{S}_{n\mid\R}$ can be decomposed as
    \begin{align*}
        \mathcal{S}_{n\mid\R}=\mathcal{S}_{n\mid\R,k\rightarrow 1}\cup\cdots\cup\mathcal{S}_{n\mid\R\cup\{k\}}\cup\cdots\cup\mathcal{S}_{n\mid\R,k\rightarrow n-s}.
    \end{align*}
    Besides, for any distinct indices $j_1,j_2\in [n]\backslash\R$, $\mathcal{S}_{n\mid\R,k\rightarrow j_1}\cap \mathcal{S}_{n\mid\R,k\rightarrow j_2}=\emptyset$. For any $j,k\in [n]\backslash\R$, we can construct a bijection between $\mathcal{S}_{n\mid\R,k\rightarrow j}$ and $\mathcal{S}_{n\mid\R\cup\{k\}}$. Let $P_{k\leftrightarrow j}$ be the permutation that swaps the $k$th and $j$th rows while leaving all other rows unchanged. Then:
    \begin{align*}
        P_{k\leftrightarrow j}\mathcal{S}_{n\mid\R,k\rightarrow j}=\mathcal{S}_{n\mid\R\cup\{k\}}.
    \end{align*}
    Moreover, for any distinct $P_{\pi_1\mid\R,k\rightarrow j},\ P_{\pi_2\mid\R,k\rightarrow j}\in \mathcal{S}_{n\mid\R,k\rightarrow j}$, we have $P_{k\leftrightarrow j}P_{\pi_1\mid\R,k\rightarrow j}\neq P_{k\leftrightarrow j}P_{\pi_2\mid\R,k\rightarrow j}$. Thus, we conclude that $P_{k\leftrightarrow j}$ is the bijection.
    
    Next, we demonstrate that for any permutation $P_{\pi\mid\R,k\rightarrow j}\in\mathcal{S}_{n\mid\R,k\rightarrow j}$, we have
    \begin{align*}
        \rank(Z-P_{\pi\mid\R,k\rightarrow j}Z)\geq \rank[Z-P_{k\leftrightarrow j}P_{\pi\mid\R,k\rightarrow j}Z].
    \end{align*}
    
     Observe that the difference between $Z-P_{\pi\mid\R,k\rightarrow j}Z$ and $Z-P_{k\leftrightarrow j}P_{\pi\mid\R,k\rightarrow j}Z$ appears only in the $k$th and $j$th rows. Without loss of generality, consider $j=1$. Denote $Z^k$ as the $k$th row of $Z$.
    \begin{align*}
        Z-P_{\pi\mid\R,k\rightarrow 1}Z=\begin{pmatrix}
            Z^1-Z^k\\
            \vdots\\
            Z^k-Z^{\pi(k)}\\
            \vdots
        \end{pmatrix},\quad Z-P_{k\leftrightarrow 1}P_{\pi\mid\R,k\rightarrow 1}Z=\begin{pmatrix}
            Z^1-Z^{\pi(k)}\\
            \vdots\\
            0\\
            \vdots
        \end{pmatrix}.
    \end{align*}
    If $\rank(Z-P_{\pi\mid\R,k\rightarrow 1}Z)< \rank(Z-P_{k\leftrightarrow 1}P_{\pi\mid\R,k\rightarrow 1}Z)$, then 
    \begin{align*}
        Z^1-Z^k, Z^k-Z^{\pi(k)}\in \sp(Z^2-Z^{\pi(2)},\dots,Z^{k-1}-Z^{\pi({k-1})},Z^{k+1}-Z^{\pi({k+1})},\dots),
    \end{align*}
    but $Z^1-Z^{\pi(k)}$ does not. This leads to a contradiction because $Z^1-Z^{\pi(k)}=(Z^1-Z^k)+(Z^k-Z^{\pi(k)})$.

    Thus we conclude that for any $k\in [n]$
    \begin{align*}
        \mu_{\R}(Z)=\frac{1}{n-s}\sum_{j=1}^{n-s} E[\rank(Z-P_{\pi\mid\R,k\rightarrow j}Z)]\geq \mu_{\R\cup\{k\}}(Z).
    \end{align*}
\end{proof}

\end{document}